\documentclass[a4paper,12pt]{article}
\usepackage{graphicx}
\usepackage{epsfig}
\begin{document}

\begin{flushright}
KLOE Memo 214\\
\today
\end{flushright}
\small

\vspace*{1.5cm}
{\LARGE \bf
\begin{center}
\boldmath
First Results from $\phi\rightarrow K_L K_S$ Decays with the KLOE Detector
\end{center}
}

\vspace{3cm}
\begin{center}
The KLOE collaboration
\end{center}
\vspace{1.5cm}

\begin{abstract}
\noindent
%%%%%osaka
The KLOE experiment has collected 2.4 $pb^{-1}$ of integrated
luminosity during the commissioning of the DA$\Phi$NE $\phi$-factory
in 1999. The performance of the detector has been studied
using the $\phi\rightarrow K_L K_S$ decays collected during this
period, yielding also  first measurements of relevant K parameters
such as masses and lifetimes.
% and main branching ratios.
A clean $K_S \rightarrow \pi^+ \pi^-$ sample is used to select
$K_L\rightarrow \pi^+ \pi^-$  CP-violating decays and $K_L \rightarrow K_S$ regeneration
events in the detector material. Results on the
regeneration probability in a beryllium-alluminum alloy and 
carbon-fiber plus aluminum composite are presented. 
%Klong identification is used
%to tag Kshort-mesons and to search for their semileptonic decays.

%%%%%lifetime
%A preliminary measurement of the $K_s \rightarrow \pi^+ \pi^-$ lifetime is presented,
%based on a limited statistics sample of 6866 decays selected from the 1999 data.
%The $K_s$ lifetime in the $\Phi$-rest frame is measured to be 5.78 $\pm$ 0.19 (syst)
%$\pm$ 0.08 (stat) ($KLOE$ $Preliminary$), which is to be compared to 5.90 $\pm$
%0.01 mm ($PDG$). This analysis is to be repeated soon on the whole 1999 $K_s$ data
%sample, after this has been reconstructed with the latest version of the offline
%package and reprocessed by the tracking monitor program in order to compute
%run-by-run the position of the luminous region and of the $\Phi$-boost in the lab system.

%%%%%regeneration
%With a sample of 434105 well reconstructed $K_{S}$ mesons produced in
%$\phi$ decays during the first period of data taking of the KLOE 
%experiment at DA$\Phi$NE we have measured the probability of 
%regeneration of 110 MeV/c $K_{L}$ mesons in the beam pipe, made of a 
%Beryllium-Aluminum alloy, and in the drift chamber wall, made  of 
%Carbon fiber. The value of the cross section and its angular 
%dependence are compared with existing theoretical predictions.

\end{abstract}

\vspace*{6truecm}
\begin{center}
{\small  Contributed paper $\sharp$ 294 to the XXX International Conference on High
  Energy Physics, Osaka 27 jul - 2 aug 2000.
}
\end{center}

\newpage

\def\ifm#1{\relax\ifmmode#1\else$#1$\fi}
\def\eps{\ifm{\epsilon}} \def\epm{\ifm{e^+e^-}}
\def\rep{\ifm{\Re(\eps'/\eps)}}  \def\imp{\ifm{\Im(\eps'/\eps)}}  
\def\DAF{DA$\Phi$NE}  \def\sig{\ifm{\sigma}}
\def\gam{\ifm{\gamma}} \def\to{\ifm{\rightarrow}}
\def\pip{\ifm{\pi^+}} \def\pim{\ifm{\pi^-}}
\def\po{\ifm{\pi^0}} 
\def\pic{\ifm{\pi^+\pi^-}} \def\pio{\ifm{\pi^0\pi^0}} 
\def\ks{\ifm{K_S}} \def\kl{\ifm{K_L}} \def\kls{\ifm{K_{L,\,S}}} 
\def\ksl{\ifm{K_S,\ K_L}} \def\ko{\ifm{K^0}}
\def\K{\ifm{K}} \def\LK{\ifm{L_K}}
\def\Kb{\ifm{\rlap{\kern.3em\raise1.9ex\hbox to.6em{\hrulefill}} K}}
\def\ab{\ifm{\sim}}  \def\x{\ifm{\times}}
\def\ff{$\phi$--factory}
\def\sta#1{\ifm{|\,#1\,\rangle}} 
\def\amp#1,#2,{\ifm{\langle#1|#2\rangle}}
\def\kob{\ifm{\Kb\vphantom{K}^0}}
\def\f{\ifm{\phi}}   \def\pb{{\bf p}}
\def\L{\ifm{{\cal L}}}  \def\R{\ifm{{\cal R}}}
\def\up#1{$^{#1}$}  \def\dn#1{$_{#1}$}
\def\etal{{\it et al.}}
\def\BR{{\rm BR}}
\def\radl{\ifm{X_0}}
\def\deg{\ifm{^\circ}} 
\def\th{\ifm{\theta}}
\def\To{\ifm{\Rightarrow}}
\def\ot{\ifm{\leftarrow}}
\def\fo{\ifm{f_0}} \def\epe{\ifm{\eps'/\eps}}
\def\pbrn{ {\rm pb}}  \def\cm{ {\rm cm}}
\def\mub{\ifm{\mu{\rm b}}} \def\s{ {\rm s}}
\def\RR{\ifm{{\cal R}^\pm/{\cal R}^0}}
\def\dt{ \ifm{{\rm d}t} } \def\dy{ {\rm d}y } \def\pbrn{ {\rm pb}}
\def\kp{\ifm{K^+}} \def\km{\ifm{K^-}}
\def\kkb{\ifm{\ko\kob}} 
\def\epe{\ifm{\eps'/\eps}}
\def\ppc{\ifm{\pi^+\pi^-}}
\def\ppo{\ifm{\pi^0\pi^0}}
\def\pppco{\ifm{\pi^+\pi^-\pi^0}}
\def\pppo{\ifm{\pi^0\pi^0\pi^0}}
\def\vare{\ifm{\varepsilon}}
\def\etap{\ifm{\eta'}}

\def\pt#1,#2,{#1\x10\up{#2}}

%\Elvpoint

%\font\bigbf=cmbx10 scaled\magstep4
%\Frontpage
%\title{\big\bf  \vspace*{-1truecm} THE KLOE COLLABORATION}
%\Elvpoint{
%\vglue -2mm

\def\B{Bari}
\def\b{\rlap{\kern.2ex\up a}}
\def\O{IHEP}
\def\o{\rlap{\kern.2ex\up b}}
\def\Fr{Frascati}
\def\fr{\rlap{\kern.2ex\up c}}
\def\Ka{Karlsruhe}
\def\ka{\rlap{\kern.2ex\up d}}
\def\Le{Lecce}
\def\le{\rlap{\kern.2ex\up e}}
\def\Mo{Moscow}
\def\mo{\rlap{\kern.2ex\up f}}
\def\N{Napoli}
\def\n{\rlap{\kern.2ex\up g}}
\def\BE{Beer-Sheva}
\def\be{\rlap{\kern.2ex\up h}}
\def\co{\rlap{\kern.2ex\up i}}
\def\Pi{Pisa}
\def\pI{\rlap{\kern.2ex\up j}}
\def\Ra{Roma I}
\def\ra{\rlap{\kern.2ex\up k}}
\def\en{\rlap{\kern.2ex\up l}}
\def\Rb{Roma II}
\def\rb{\rlap{\kern.2ex\up m}$\,$}
\def\Rc{Roma III}
\def\rc{\rlap{\kern.2ex\up n}}
\def\su{\rlap{\kern.2ex\up o}}
\def\T{Trieste/Udine}
\def\t{\rlap{\kern.2ex\up q}}
\def\V{Virginia}
\def\v{\rlap{\kern.2ex\up r}}
\def\Z{Associate member}
\def\hsa{ \ }

%\author{ 
%{\baselineskip 12pt
%\parskip=0pt
%\parindent=0pt}
\normalsize\noindent
M.~Adinolfi\rb,\hsa 
A.~Aloisio\n,\hsa
F.~Ambrosino\n,\hsa
A.~Andryakov\mo,\hsa
A.~Antonelli\fr,\hsa %\\[-1.5mm] \normalsize
M.~Antonelli\fr,\hsa 
F.~Anulli\fr,\hsa
C.~Bacci\rc,\hsa
A.~Bankamp\ka,\hsa
G.~Barbiellini\t,\hsa
% ...
%}  %%%%

%\maketitle

\vglue 2mm

\def\aff#1{Dipartimento di Fisica dell'Universit\`a e Sezione INFN, #1, Italy.}

%{\baselineskip=12pt
%\parskip=0pt
%\parindent=0pt
\def\hsb{\hskip 2.8mm}

\leftline{\b\hsb \aff{\B}}
\leftline{\o\hsb Institute of High Energy Physics of Academica Sinica, 
Beijing, China.}
\leftline{\fr\hsb  Laboratori Nazionali di Frascati dell'INFN, Frascati, Italy.}
\leftline{\ka\hsb  Institut f\"ur Experimentelle Kernphysik, Universit\"at \Ka,
Germany.}
\leftline{\le\hsb \aff{\Le}}
\leftline{\mo\hsb Institute for Theoretical and Experimental Physics, Moscow,
Russia.}
\leftline{\n\hsb Dipartimento di Scienze Fisiche dell'Universit\`a e 
Sezione INFN, \N, Italy.}
\leftline{\be\hsb Physics Department, Ben-Gurion University of the Negev,
Israel.}
\leftline{\co\hsb Physics Department, Columbia University, New York, USA.}
\leftline{\pI\hsb \aff{\Pi}}
\leftline{\ra\hsb \aff{\Ra}}
\leftline{\en\hsb Dipartimento di Energetica dell'Universit\`a, Roma I, Italy.}
\leftline{\rb\hsb \aff{\Rb}}
\leftline{\rc\hsb \aff{\Rc}}
\leftline{\su\hsb Physics Department, State University of New York 
at Stony Brook, USA.}
\leftline{\t\hsb \aff{\T}}
\leftline{\v\hsb Physics Department, University of Virginia, USA.}
\leftline{\rlap{\kern.2ex\up*}\hsb\Z}    
%  }  

\newpage

\section{Introduction}
The DA$\Phi$NE $\phi$ factory \cite{DAFNE} has come into operation in july 1999,
delivering 2.4 $pb^{-1}$ of integrated luminosity to the KLOE experiment
during its commissioning period. Such data allowed to perform a careful
check of the detector operation and performances \cite{BERTOLUCCI}.\par
A sample of $\phi\rightarrow K_SK_L$ decays provided 
a source of monochromatic, 110 MeV/c $(\beta=0.21)$, 
beams of $K_{S}$ and $K_{L}$ mesons, which were reconstructed in various
decay modes.\par
A preliminary measurement of the $K_S$ lifetime was performed in a
subsample of events. Charged decays oh the $K_L$ were also reconstructed,
and a sample of CP violating $K_L\rightarrow \pi^+\pi^-$ decays was selected. 
The probability of $K_{L} \to K_{S}$ regeneration in the 
materials of the KLOE detector was also measured. With the available
statistics it was possible
to clearly identify regeneration events produced in the beam-pipe, made 
of a Beryllium-Aluminum alloy, and in the drift chamber wall, made 
of Carbon fiber, and to measure the angular dependence of the cross 
section.

\section{Experimental setup}

The KLOE experiment \cite{BERTOLUCCI} consists of a large volume 
tracking detector surrounded by a hermetic calorimeter both immersed 
in the axial magnetic field of 0.56 T produced by a superconducting coil. 
The tracking detector is a cylindrical drift chamber with alternated 
stereo views. The sensitive volume, 25 cm inner radius, 195 cm outer radius 
and 320 cm length is filled with a 90\%Helium-10\%Isobutane gas mixture.
Charged particle tracks are reconstructed in 12582 drift cells arranged 
in 58 concentric layers. The spatial resolution in the transverse 
plane is about 150 $\mu$m.\par
The calorimeter, made with 1 mm diameter
scintillating fibers immersed in 0.5 mm grooved lead plates, has a thickness 
of 23 cm corresponding to 15 $X_{0}$. It is segmented in 
288 sectors in the barrel and 200 in the endcaps, each sector being 
subdivided in five longitudinal samplings. Energy clusters are 
reconstructed with a resolution $\sigma_{E}/E = 0.057/ [E \
(GeV)]^{1/2}\oplus 0.006$
and a space resolution of few cm. Time of flight measurement
with a resolution of 
$\sigma_{t}  = 0.05 \ ns / [E \ (GeV)]^{1/2} \oplus 0.14 \ ns $
is used for particle identification.\par
Data were taken at the energy of the $\phi$ resonance for six weeks 
in the period august-december 1999 during the commissiong 
of the DA$\Phi$NE collider. Typical values of the beam 
parameters are listed in Table \ref{TABLE1}. The beams cross with a
period multiple of 2.7 ns and with an angle of 25 mrad in the horizontal plane. 
This implies a small boost of 13 MeV of the center of mass relative to 
the experiment.

\begin{table}[htb]
\centering
\begin{tabular}{| c | c |}
\hline
 numer of bunches & 20 $e^{+}$ + 20 $e^{-}$ \\
 beam currents & 300 mA \\
 beam luminosity & $10^{30} \ cm^{-2} s^{-1}$ \\
 luminosity lifetime & $\approx$ 30 min \\
\hline
\end{tabular}
\caption{Typical values of the DA$\Phi$NE beam parameters during data dating.}
\protect\label{TABLE1}
\end{table}

The trigger requirement during most of the data taking was  at least two calorimeter energy deposits in the configurations Barrel-Barrel or
 Barrel-EndCap
with E(Barrel)$ > $ 50 MeV and E(Endcap)$ > 90$ MeV.
%\begin{itemize}
%\item at least two calorimeter energy deposits of more than 30??? MeV 
%in the configurations Barrel-Barrel or Barrel-EndCap;
%\item ???
%\end{itemize}
This trigger has an average efficiency \cite{PALUTAN} of about 90\% for
most of the $\phi$ decays and of about 84\% for $\phi \to K_{L}K_{S}$ 
events with $K_{L}$ and $K_{S}$ decaying into charged particles of 
interest for this analysis. The trigger rate was typically 1.5 kHz 
mostly due to cosmic rays.
The luminosity was measured with an accuracy of about 5\% by recording 
Bhabha scattering events in the polar angle interval 
$22^{o} < \theta < 158^{o}$.

\subsection{Data sample}
About 8 million $\phi$ decays were collected during data taking.
The data were filtered against cosmics and machine background, and then
classified in 5 different classes \cite{EVCL}:
\begin{itemize}
\item[1.] $\phi\rightarrow K_SK_L$
\item[2.] $\phi\rightarrow K^+K^-$
\item[3.] $\phi$ radiative decays
\item[4.] $\rho\pi$ events
\item[5.] Bhabhas
\end{itemize}
This analysis used only events in class 1 where a candidate $K_S$ charged
decay was present, satisfying the following requirements:
one vertex made with two tracks of opposite curvature 
with $\rho = (x^{2} + y^{2})^{1/2} < 4$ cm, $| z | < 8$ cm, 
two pion invariant mass $400 < m < 600$ MeV, and total momentum 
$50 < p < 120$ MeV.\par

\section{Measurement of the $K_S$ lifetime}

\subsection{Measurement Technique}

The position of the secondary vertex (SV) of $\pi^+$$\pi^-$ pairs from $K_s$
decays, together with the knowledge of the position of $e^+$$e^-$ primary
vertex (PV, the luminous region) allows the measurement of the $K_s$
decay length in the $\phi$ frame, $\lambda_s=\beta \gamma c\tau_s $.
$X_{PV}$ and $Y_{PV}$ are reconstructed run-by-run from Bhabha scattering events with a
typical accuracy of few tens of microns and have widths L(X), L(Y) which are small
compared to the 6 mm $K_s$ decay length \cite{DAF99}, \cite{TRKMON}. L(X) = 1.0 mm
is measured by KLOE and L(Y) = few tens of microns is measured by DA$\Phi$NE.
$Z_{PV}$ is also reconstructed run-by-run
with 100-200 $\mu$m accuracy, but it has a width of 12-14 mm, which
distorts the $K_s$ lifetime distributions. Therefore, $Z_{PV}$ is computed event
by event using the polar angle ($\theta_s$) of the $K_s$ momentum ($p_s$) :
\begin{equation}
Z_{PV}=Z_{SV}+cot(\theta_s)\times\sqrt{(X_{PV}-X_{SV})^2+(Y_{PV}-Y_{SV})^2}.
\end{equation}

The effectiveness of this procedure is demonstrated by extrapolating the $K_s$
trajectory from the SV to the PV along ${\vec{p_s} / p_s}$. 
Fig.\ref{intercept} shows the difference between the intercept and
the PV coordinate for x, y, and z, respectively (the selection of the data sample
in these plots is described in the following section).\par

%%%%%%%%%%%%%%%%%%%%%%%%%%
\begin{figure}[htpb]
\begin{center}
%\vspace{-4truecm}
\epsfig{file=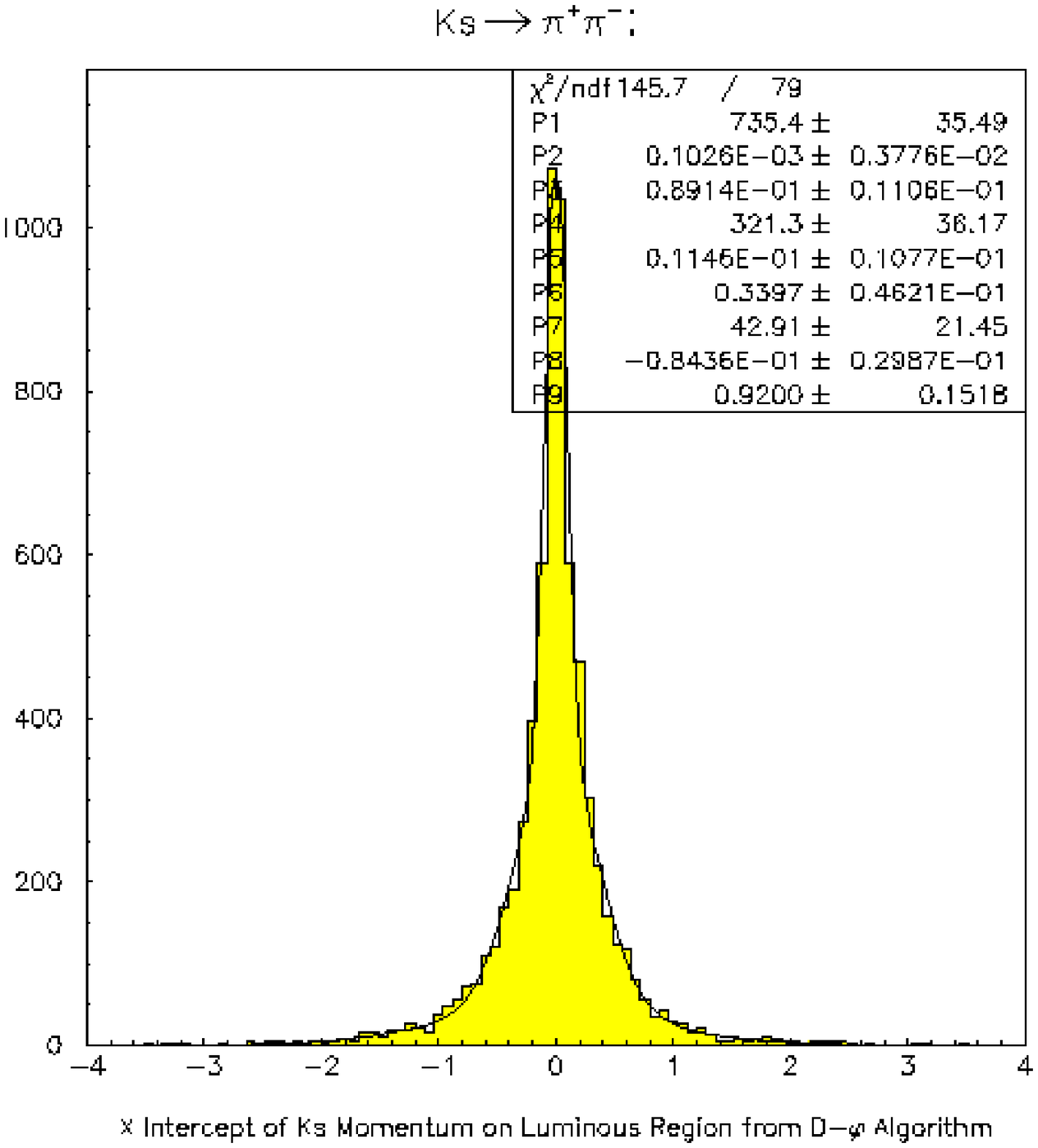,width=7.3cm}
\epsfig{file=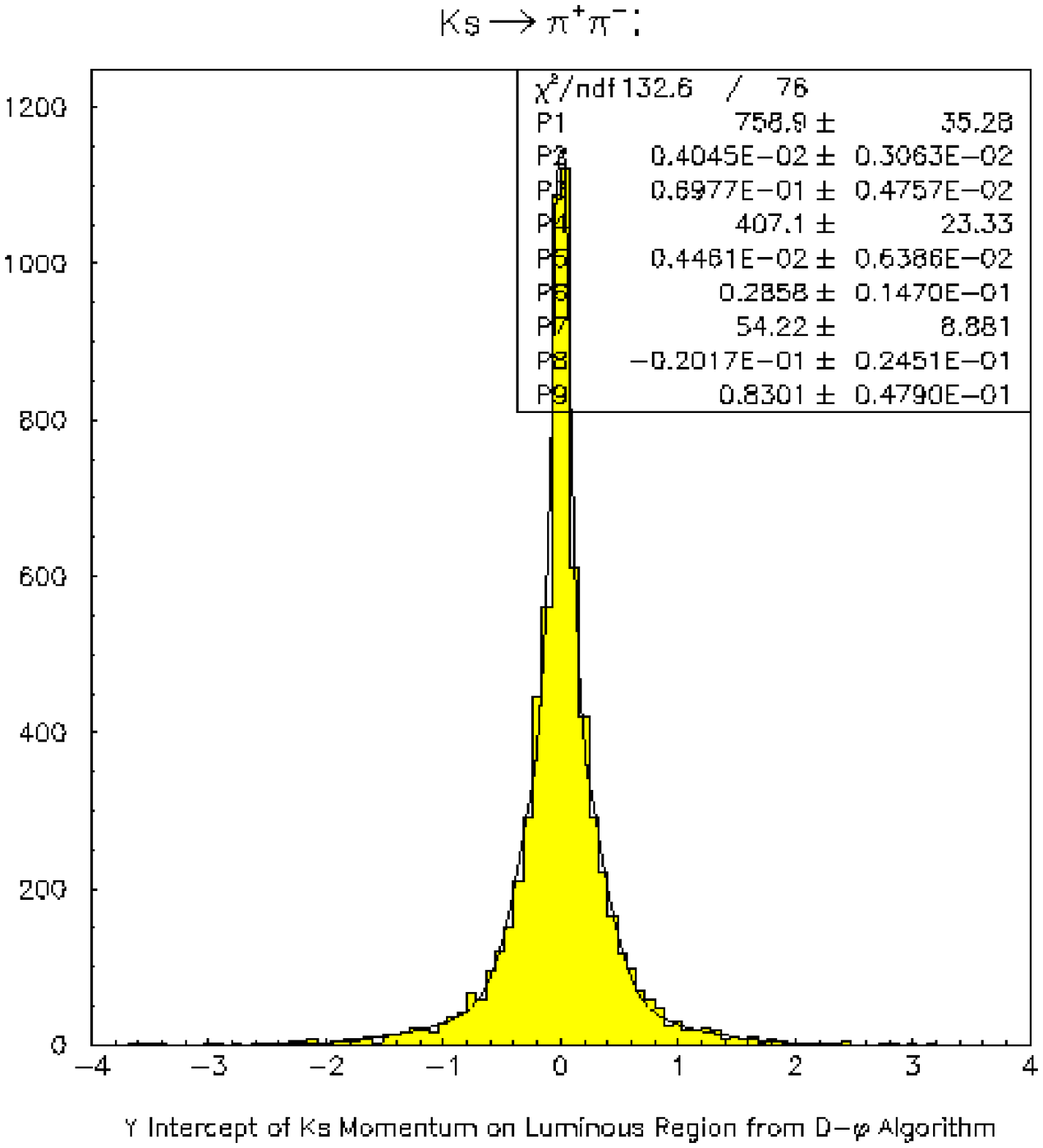,width=7.3cm}
\epsfig{file=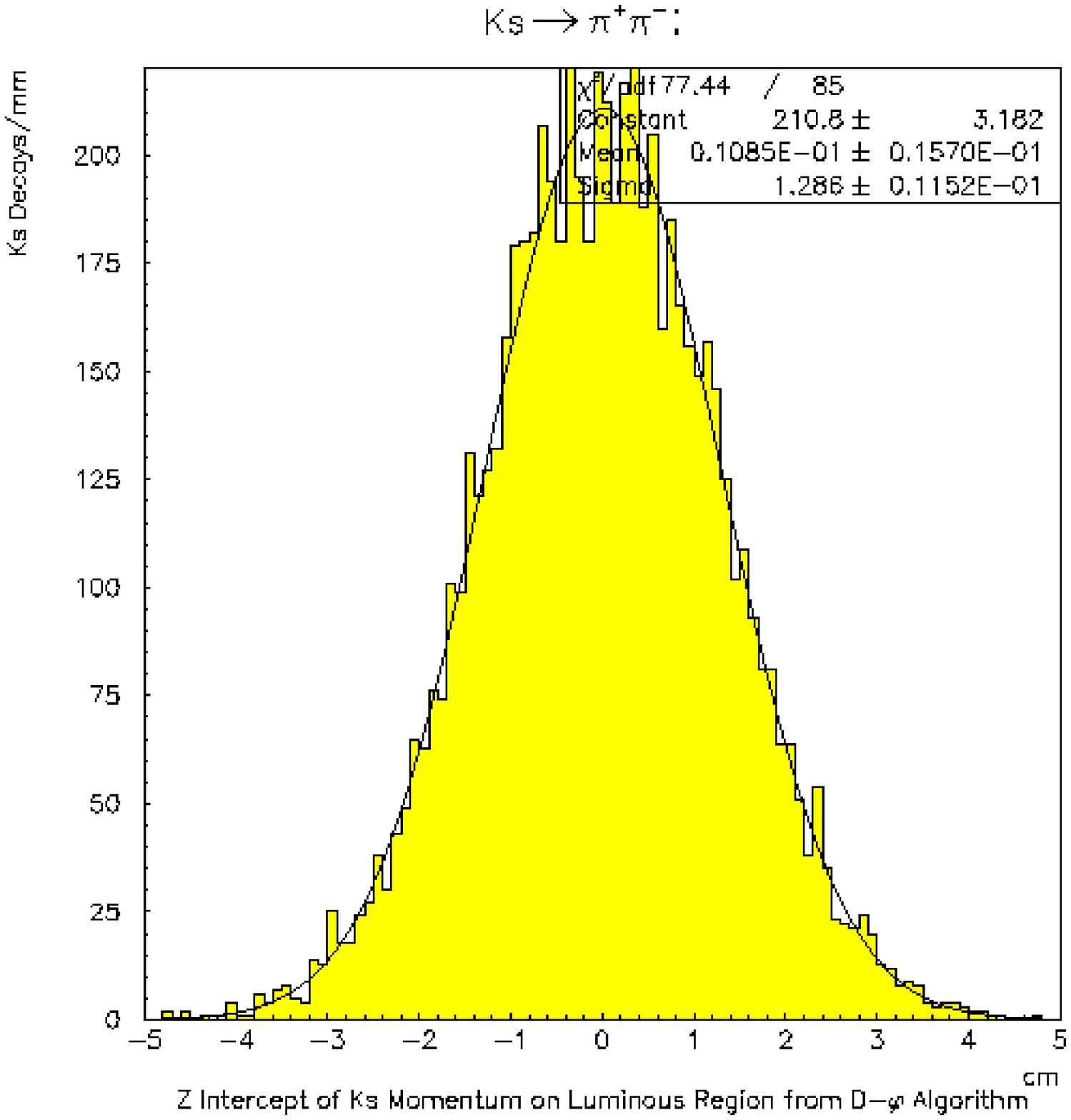,width=7.3cm}
%\vspace{-2truecm}
\caption[]{Projections of  $\vec{p}_s$ distance of closest approach
to the PV}
\label{intercept}
\end{center}
\end{figure}
%%%%%%%%%%%%%%%%%%%%%%%
To avoid the systematic distortion of the lifetime distribution due to the SV
resolution (which from Figs.\ref{intercept} is shown to
be comparable to the $K_s$ lifetime itself), the following estimator is used:
\begin{equation}
\lambda=(\vec{r}_{SV}-\vec{r}_{PV})\cdot{\vec{p_s} \over p_s}.
\end{equation}
$\lambda$ is the $K_s$  decay length projected on the $\vec{p_s}$
direction. $\lambda_s$ can
then be extracted from the fit to the $\lambda$ distribution using an
exponential appropriately smeared with PV and SV resolution functions. 
Since from direct measurement L(X) = 1.0 mm and from Monte Carlo studies
$\sigma(\theta_s), \sigma(\phi_s) \sim$ few tens of mrad, the
resolution on the plots of Fig.\ref{intercept} can be interpreted in terms 
of the SV resolution, that turns out to be by far the dominant contribution.\par

Finally, $\lambda$ is Lorentz-transformed to the $\phi$ system
using the run-average $\phi$ boost as measured
from Bhabha events \cite{TRKMON}. 

%%%%%%%%%%%%%%%%%%%%%%%%%%
%\begin{figure}[hb]
%\vspace{-0.3in}
%\begin{center}
%\leavevmode
%\epsfxsize \textwidth
%\epsfysize 15.cm
%\vspace{0.3in}
%\epsffile{lvsphi_cmslab.ps}
%\vspace{-0.in}
%\caption[]{$\lambda$ vs $\phi_s$ in the LAB and CMS systems.}
%\label{lvsphi_cmslab}
%\end{center}
%\end{figure}
%%%%%%%%%%%%%%%%%%%%%%%

\subsection{Sample Selection}

This analysis was performed using only a subsample of events  
for which the position of the luminous region had been computed after data 
taking. 
The selection required the simultaneous presence of the $K_S$ candidate
vertex and of a second two-track vertex outside a sphere of 11 cm radius. 
If there is a third vertex
the event is rejected. The following cuts are then imposed to the $K_s$ 
candidates inside the beam pipe:

\begin{itemize}
\item[(1)] 493 MeV $< M_s <$ 497 MeV,
\item[(2)] in the $\phi$-CMS frame, 105 MeV $< P_s <$ 114 MeV,
\item[(3)] in the $\phi$-CMS frame, 45$^{\circ} < \theta_s < 135^{\circ}$,
\item[(4)] $|cot\theta(\pi,LAB)|< 1.0$ for both pions.
\end{itemize}
The final selected sample is 6866 decays.\par

%%%%%%%%%%%%%%%%%%%%%%%%%%
%\begin{figure}[htpb]
%\vspace{-0.3in}
%\begin{center}
%\epsfig{file=cutsimon.eps,width=12cm,height=12cm}
%\vspace{-1.in}
%\caption[]{$K_s$ sample selection cuts}
%\label{lcuts}
%\end{center}
%\end{figure}
%%%%%%%%%%%%%%%%%%%%%%%

\subsection{Results}

The lifetime distribution for the 6866 decays was fit to four different functions,
with the constraint of the total number of decays:
\begin{itemize}
\item[(1)]  fit to an exponential smeared with two gaussian resolutions.
\item[(2)]  fit to an exponential smeared with three gaussian
resolutions. 
\item[(3)]  fit to an exponential smeared with two gaussian resolutions
plus a third gaussian modeling a zero-lifetime component (due to background and/or
to any other systematic effect). 
\item[(4)]  fit to an exponential smeared with three gaussian
resolutions plus a fourth gaussian modeling a zero-lifetime component (due to
background and/or to any other systematic effect). 
\end{itemize}
As an exanple cases (2) and (4) are shown in fig.\ref{lifefit}.
In both fits the parameter P1
is $\lambda_s$ and P2 and P3 (P4 and P5) are the populations and standard 
deviation of the first (second) smearing gaussian.
In case (2) P6 is the standard deviation of the third smearing gaussian, 
while in case (4) P6 and P7 are the population and standard deviation of
the third smearing gaussian and P8 is the standard deviation of
the fourth non-smearing gaussian. The results of the four fits are
reported in table \ref{TABLE3}.\par

\begin{table}[htb]
\centering
\begin{tabular}{| c | c |}
\hline
fit function & $\lambda_s$ (mm) \\
\hline
2g smearing & 5.71 $\pm$ 0.06 \\
3g smearing & 5.71 $\pm$ 0.06 \\
2g smearing + 1g non smearing & 5.90 $\pm$ 0.08 \\
3g smearing + 1g non smearing & 5.83 $\pm$ 0.07 \\
\hline
\end{tabular}
\caption{Typical values of the DA$\Phi$NE beam parameters during data dating.}
\protect\label{TABLE3}
\end{table}

The average $\lambda_s$ value from the four fits is 5.78 mm. The fit statistical
error on $\lambda_s$ is taken as the largest fit error, 0.08 mm, while the fit
systematic error is taken as half the difference between the maximum and 
minimum values of $\lambda_s$, 0.10 mm. The preliminary result is
then:
$$ \lambda_s\, =\, 5.78\, \pm\, 0.08\, (stat)\, \pm\, 0.10\, (syst)\, mm$$

%%%%%%%%%%%%%%%%%%%%%%%%%%
\begin{figure}[htpb]
\vspace{-3truecm}
\begin{center}
\epsfig{file=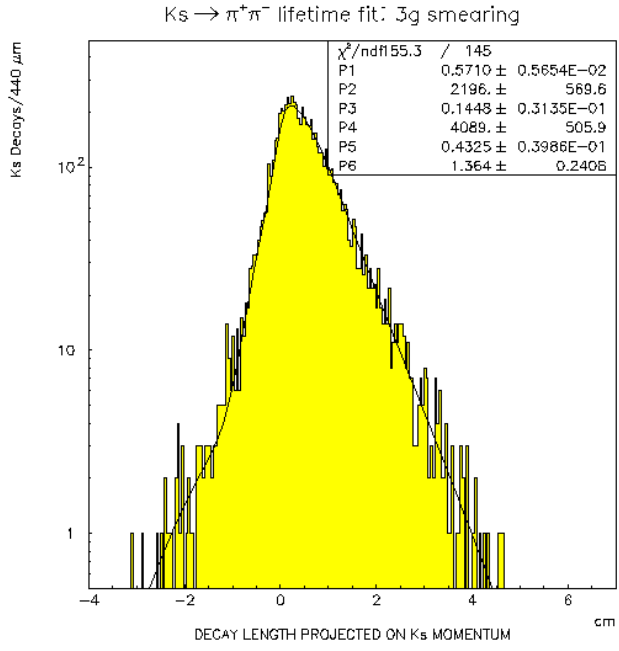,width=10cm}
%\epsfig{file=ks_2g1g.eps,width=10cm}
%\vspace*{-5truecm}
\epsfig{file=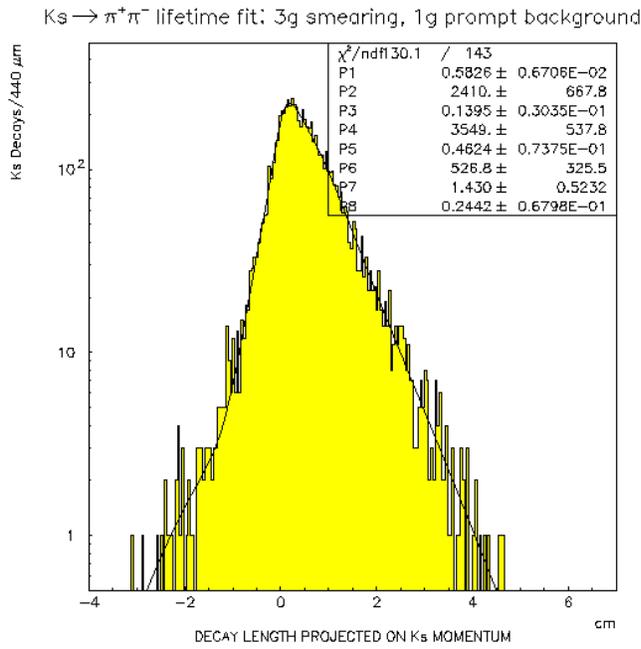,width=10cm}
%\vspace{-1.in}
\caption[]{$K_s$ decay length fit to an exponential smeared with three
  gaussians without and with a fourth non-smearing gaussian.}
\label{lifefit}
\end{center}
\end{figure}
%%%%%%%%%%%%%%%%%%%%%%%

%%%%%%%%%%%%%%%%%%%%%%%%%%
%\begin{figure}[htpb]
%\vspace{-0.3in}
%\begin{center}
%\leavevmode
%\epsfxsize \textwidth
%\epsffile{lvsphi_bpos.ps}
%\vspace{-1.in}
%\caption[]{$\lambda$ vs $\phi_s$ for november data with the correct position
%of the luminous region, $\mu$(X) = 4.5 mm, $\mu$(Y) = 1.0 mm.}
%\label{lvsphi_bpos}
%\end{center}
%\end{figure}
%%%%%%%%%%%%%%%%%%%%%%%

%%%%%%%%%%%%%%%%%%%%%%%%%%
%\begin{figure}[htpb]
%\vspace{-0.3in}
%\begin{center}
%\leavevmode
%\epsfxsize \textwidth
%\epsffile{lvsphi_nobpos.ps}
%\vspace{-1.in}
%\caption[]{$\lambda$ vs $\phi_s$ for november data with the incorrect position
%of the luminous region $\mu$(X) = $\mu$(Y) = 0. mm (instead of $\mu$(X)=4.5 mm, 
%$\mu$(Y)=1.0 mm).}
%\label{lvsphi_nobpos}
%\end{center}
%\end{figure}
%%%%%%%%%%%%%%%%%%%%%%%

\section{Measurement of the $K_L\rightarrow K_S$ regeneration cross section}
 
Due to the different absorbtion cross section of 
$K^{o}$ and $\bar{K}^{o}$, a pure beam of $K_{L}$ mesons will regenerate 
\cite{PAIS} $K_{S}$ mesons when traversing the detector. 
Thus $K_{L} \to K_{S} \to \pi \pi$ is a potential source of 
background for the rare $K_{L} \to \pi \pi$ $CP$-violating decays in 
experiments where high precision in event counting is required.

Regeneration is well studied at high energy but there is lack of  
experimental results at energies below 500 MeV. On the other hand
theoretical predictions on the $K_{L} \to K_{S}$ regeneration cross 
section at low energy suffer of large uncertainties. It has been shown  
\cite{KLEINKNECHT} that for a thin regenerator, 
$t < \lambda_{S}$, the coherent regeneration is negligible. 
Inelastic regeneration can also be considered negligible in 110 MeV/c 
$K_{L}$-nuclei interactions.  
Theoretical calculations \cite{BALDINI} on incoherent $K_{L}$ regeneration, 
based on the eikonal approximation and the 
Woods-Saxon form for the nuclear potential, foresee values in 
the range of 20-50 mb for the regeneration cross section on nuclei.

\subsection{Data analysis}
$K_{S} \to \pi^{+} \pi^{-}$ candidates were selected as described in
sect. 2.1. The additional requirement of the presence of at least one more 
vertex, with two unlike sign tracks, was made.

The distribution of the invariant mass and of the vector sum of the momenta 
in the $\phi$-reference system are shown in Fig.\ref{KSM&P} for the 
selected 604226 vertices (592687 events). 
A clear peak at the $K_{S}$ mass and at $p_{S} = 110$ MeV/c is observed. 
The off-peak values are mainly due to $\pi \to \mu \nu$ decays in flight and 
to reconstruction errors.

\begin{figure}
\begin{center}
\epsfig{file=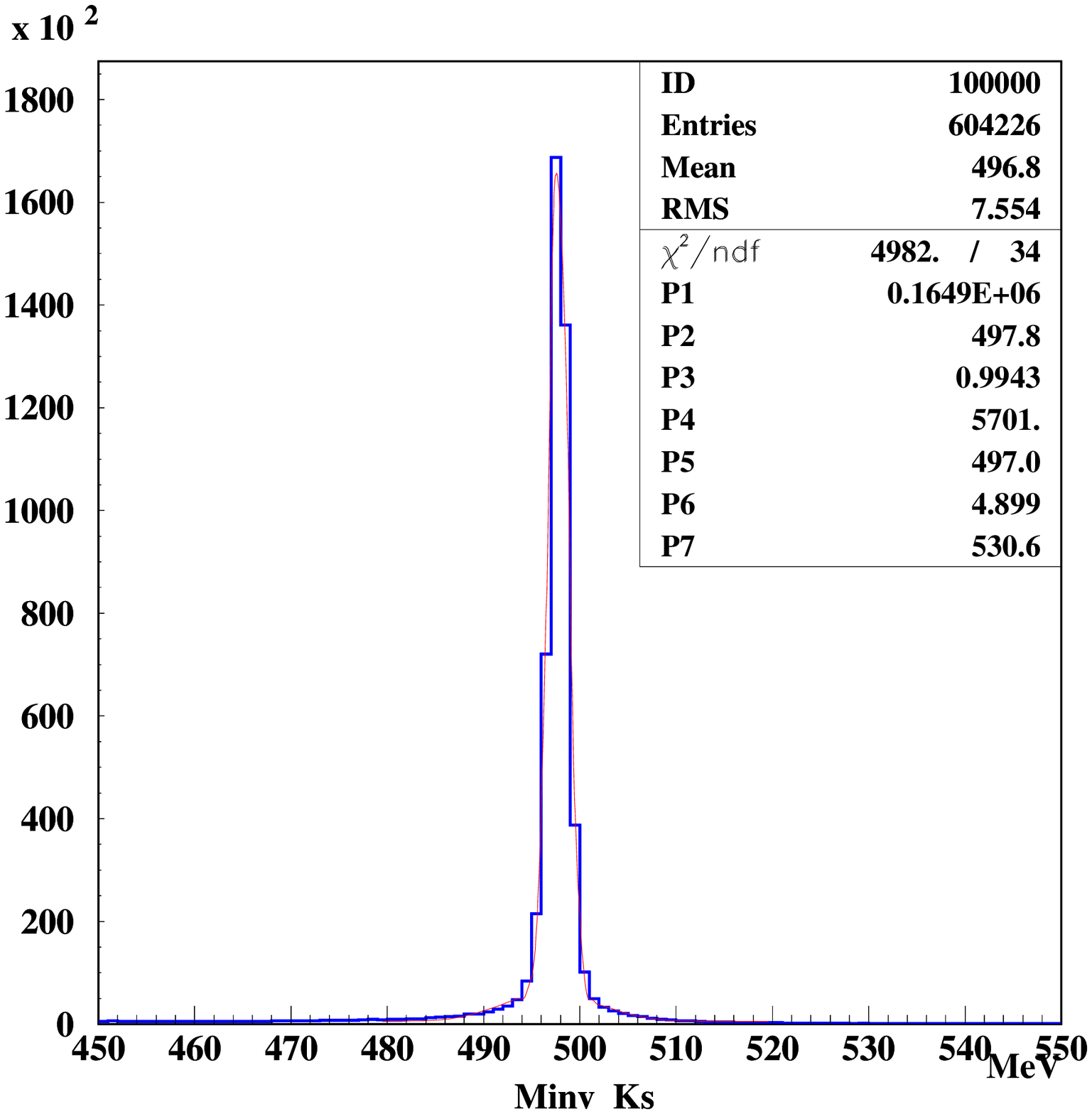,width=7cm}
\epsfig{file=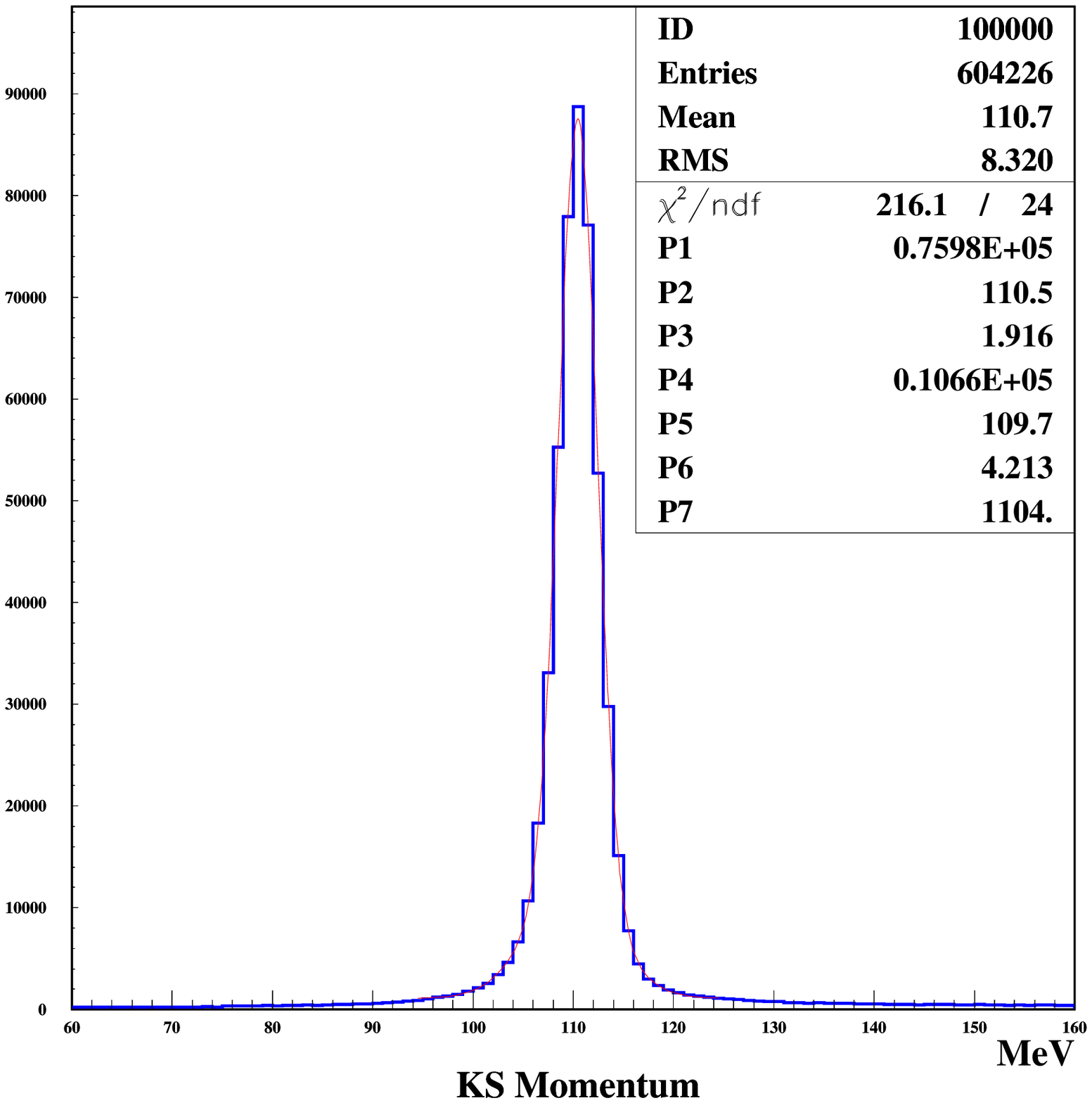,width=7cm}
\caption{Distribution of the $K_{S}$ vertex invariant mass and momentum}
\label{KSM&P}
\end{center}
\end{figure}

%\begin{figure}[htb]
%       \centering
%       \includegraphics[width = 3cm]{kloefigure.eps}
%       \caption{Distribution of the $K_{S}$ vertex invariant mass and 
%       momentum}
%       \label{KSM&P}
%\end{figure}

Fitting the distributions with two gaussians, the peak r.m.s. widths 
are $\sigma_{m} =  1.0 $ and $\sigma_{p} = 1.9 $ MeV. 
445347 well reconstructed $K_{S} \to \pi^{+} \pi^{-}$ decays were 
selected requiring
\[ \left( \frac{m - \langle m \rangle}{\sigma_{m}} \right)^{2} +   
   \left( \frac{p - \langle p \rangle}{\sigma_{p}} \right)^{2} < 16 \]
To search for the associated $K_{L}$ decay vertex we first defined the 
$K_{L}$ origin (the $\phi$ vertex) as the distance of closest approach 
between the $K_{S}$ direction and the beam line ($z$ axis).  
Events were retained requiring for the vertex position 
\[ \left( \frac{x - \langle x \rangle}{\sigma_{x}} \right)^{2} + 
   \left( \frac{y - \langle y \rangle}{\sigma_{y}} \right)^{2} +
   \left( \frac{z - \langle z \rangle}{\sigma_{z}} \right)^{2} < 9 \]
where $\sigma_x =0.63\, cm$, $\sigma_y = 0.59\, cm$ and $\sigma_z = 1.38\,
   cm$ are the r.m.s. widths of the $\phi$ vertex distributions.
433786 vertices (433760 events) satisfied this cut.
When more than one vertex were found in the $K_{S}$ fiducial volume, 
the vertex associated with the invariant mass closer to the peak value
was retained. This is the $K_{S} K_{L}$ sample. 

%\begin{figure}
%\begin{center}
%\epsfig{file=phivertex.eps,width=7cm}
%\caption{Distribution of the $\phi$-vertex coordinates}
%\label{PHIVERTEX}
%\end{center}
%\end{figure}

%\begin{figure}[htb]
%       \centering
%       \includegraphics[width = 3cm]{kloefigure.eps}
%       \caption{Distribution of the $\phi$-vertex coordinates}
%       \label{PHIVERTEX}
%\end{figure}

%A fit with an exponential and two gaussians to the distribution of the $K_{S}$ decay distance 
%(Fig.\ref{KSVERTEX}) gives $\lambda = 0.577 \ \pm 0.010$ cm, in  
%agreement with the expected value of 0.582 cm, and a vertex resolution 
%of 0.7 cm.

%\begin{figure}
%\begin{center}
%\epsfig{file=ksdecay.eps,width=7cm}
%\caption{Distribution of the $K_{S}$ decay length in the $\phi$ refrence system}
%\label{KSVERTEX}
%\end{center}
%\end{figure}

%\begin{figure}[htb]
%       \centering
%       \includegraphics[width = 3cm]{kloefigure.eps}
%       \caption{Distribution of the $K_{S}$ decay length in the $\phi$ 
%       reference system}
%       \label{KSVERTEX}
%\end{figure}

Associated charged decays of the $K_{L}$ were then searched for. The $K_{L}$ 
decay vertex is any second vertex reconstructed with two unlike sign tracks 
found in a cone opposite to the $K_{S}$ direction in the $\phi$ reference 
system. Fig.\ref{KSKLANGLE} shows the distribution of the angle 
$\delta$ between the direction of the $K_{L}$, defined by $- \vec{p}_{S}$, 
and the line joining the $K_{L}$ vertex and the $\phi$ vertex, 
for different intervals of the distance $d$ between the two vertices.
From such distributions we derived the angle resolution, 
$\sigma_{\delta} =  18$ mrad, and the transverse vertex resolution, 
$\delta r_{\perp} =  0.56 $ cm. The $K_{L}$ vertex was accepted if
\[ \delta < 4 \ \left[ {\sigma_{\delta}}^{2} + 
   \left( \frac{\delta r_{\perp}}{ d } \right)^{2} \right]^{1/2} \]
   
\begin{figure}
\begin{center}
\epsfig{file=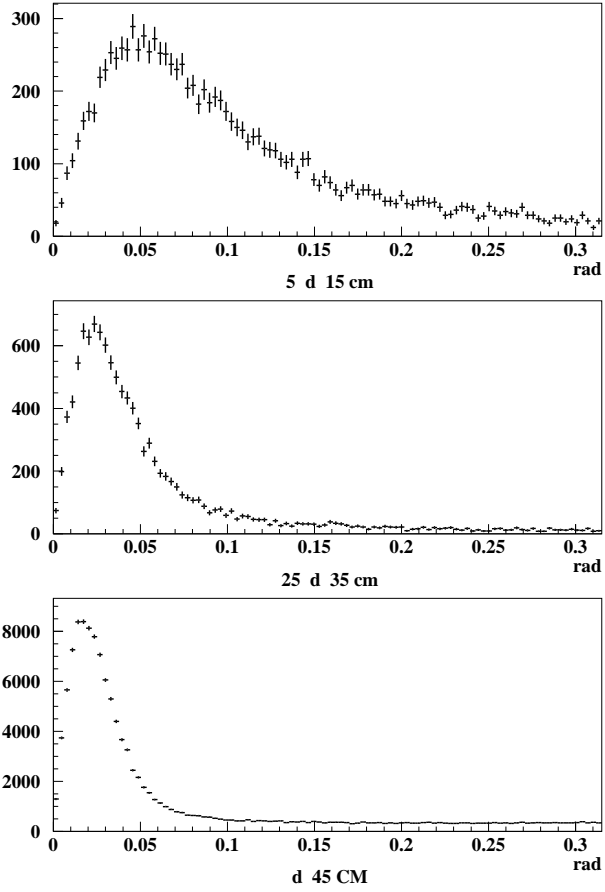,width=9cm}
\caption{Distribution of the angle between the $K_{L}$ direction and the line joining the $\phi$-vertex and the $K_{L}$-vertex for $ 5 < r < 15 $ cm, $ 25 < r < 35 $ cm, $ r > 45 $ cm.}
\label{KSKLANGLE}
\end{center}
\end{figure}

%\begin{figure}[htb]
%       \centering
%       \includegraphics[width = 3cm]{kloefigure.eps}
%       \caption{Distribution of the angle between the $K_{L}$ direction and 
%       the line joining the $\phi$-vertex and the $K_{L}$-vertex}
%       \label{KSKLANGLE}
%\end{figure}

If more than one $K_{L}$ vertex were found (only in 0.15\% of the events) 
we selected the vertex with the smaller angle. This cut selected 
134997 events.

Different decays show different behaviour in terms of the missing momentum
and the associated missing mass 
computed assuming the pion mass for all charged particles. 
Fig.\ref{KLPMMISS} shows the correlation of $p_{mis}$ and $M^{2}_{mis}$: 
$K_{L} \to \pi^{+} \pi^{-} \pi^{o}$ decays populate the region of 
$M^{2}_{mis} = m^{2}_{\pi^{o}}$; $K_{L} \to \pi \ell \nu$ decays are 
clearly separated in two bands with $M^{2}_{mis} < 0$; 
$K_{L} \to \pi^{+} \pi^{-}$ decays are peaked around $p_{mis} = 0$, 
$M^{2}_{mis} = 0$; $K_{L} \to K_{S} \to \pi^{+} \pi^{-}$ 
"elastic" events are expected to populate the band with 
$p_{mis} =  (- M^{2}_{mis})^{1/2}$.

Fig.\ref{KLDISTANCE} shows the distribution of the $K_{L}$ decay length. 
Fitting the distribution with an exponential in the region  
40 cm $< r < $ 150 cm, we obtain $\lambda = 333 \pm 13$ cm 
for the average $K_{L}$ decay length in good agreement with the expected 
value of 343 cm, thus indicating a reconstruction efficiency constant over the range of the fit.

\begin{figure}[htb]
\begin{center}
\epsfig{file=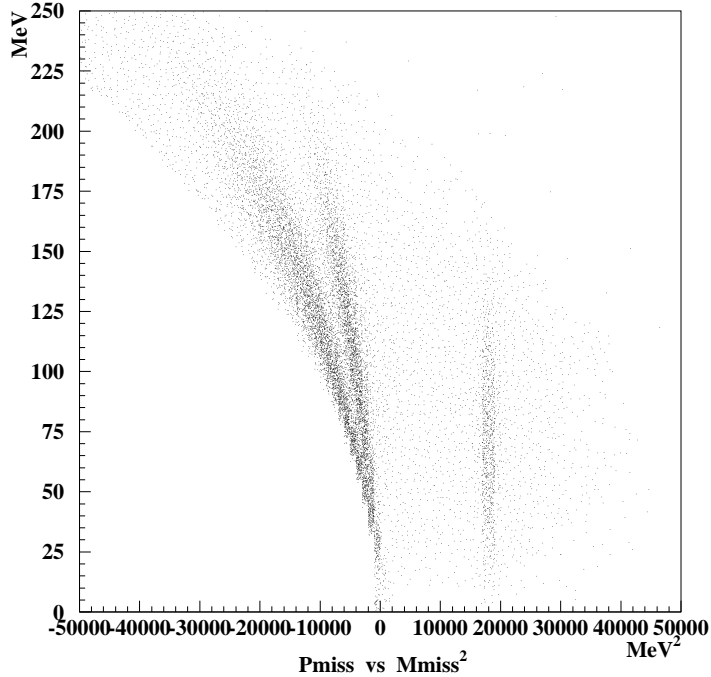,width=10cm}
       \caption{Correlation between the missing momentum and the squared 
       missing mass for charged decays of the $K_{L}$}
       \label{KLPMMISS}
\end{center}
\end{figure}

\begin{figure}
\begin{center}
\epsfig{file=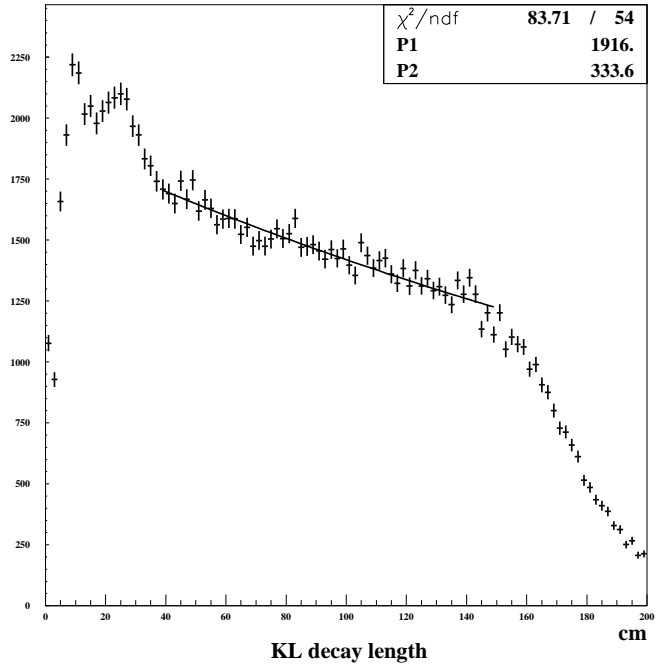,width=10cm}
\caption{Distribution of the $K_{L}$ decay length}
\label{KLDISTANCE}
\end{center}
\end{figure}

%\begin{figure}[htb]
%       \centering
%       \includegraphics[width = 3cm]{kloefigure.eps}
%       \caption{Distribution of the $K_{L}$ decay length}
%       \label{KLDISTANCE}
%\end{figure}

To select $K_{L} \to \pi^{+} \pi^{-}$ decays and 
$K_{L} \to K_{S} \to \pi^{+} \pi^{-}$ elastic events we required the 
invariant mass associated with the $K_{L}$ vertex to be the neutral kaon
mass. Fig.\ref{KLINVMASS} shows the invariant mass distribution fitted with 
a polynomial and a gaussian of r.m.s. width $\sigma_{m} =  1.1 $ MeV. 
The distribution is divided in a signal band, $m = 497.7 \pm 4$ MeV, 
with 1991 events and two side-bands, 
$m = 491.7 \pm 2$ MeV and $m = 503.7 \pm 2$ MeV with a total of
1534 events populated mostly by $K_{L}$ semileptonic decays. 

\begin{figure}
\begin{center}
\epsfig{file=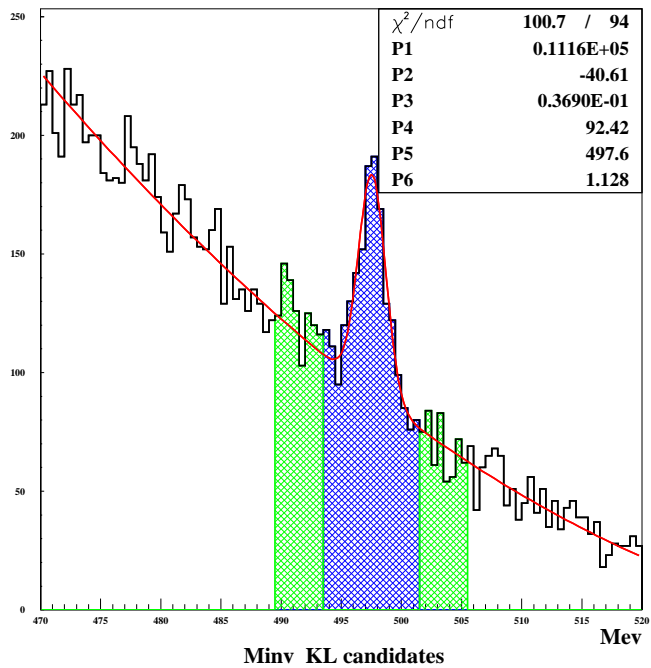,width=10cm}
\caption{Distribution of the invariant mass for charged decays of the $K_{L}$}
\label{KLINVMASS}
\end{center}
\end{figure}

%\begin{figure}[htb]
%       \centering
%       \includegraphics[width = 3cm]{kloefigure.eps}
%       \caption{Distribution of the invariant mass for charged decays of the $K_{L}$}
%       \label{KLINVMASS}
%\end{figure}

The distribution of the decay distance is shown in Fig.\ref{KL2PDISTANCE} 
for the signal band and the side-bands: the two peaks around 
$r = 10$ cm and $r = 28$ cm are interpreted as due to 
$K_{L} \to K_{S}$ regeneration in the spherical beam pipe and in the 
cylindrical drift chamber inner wall.

\begin{figure}
\begin{center}
\epsfig{file=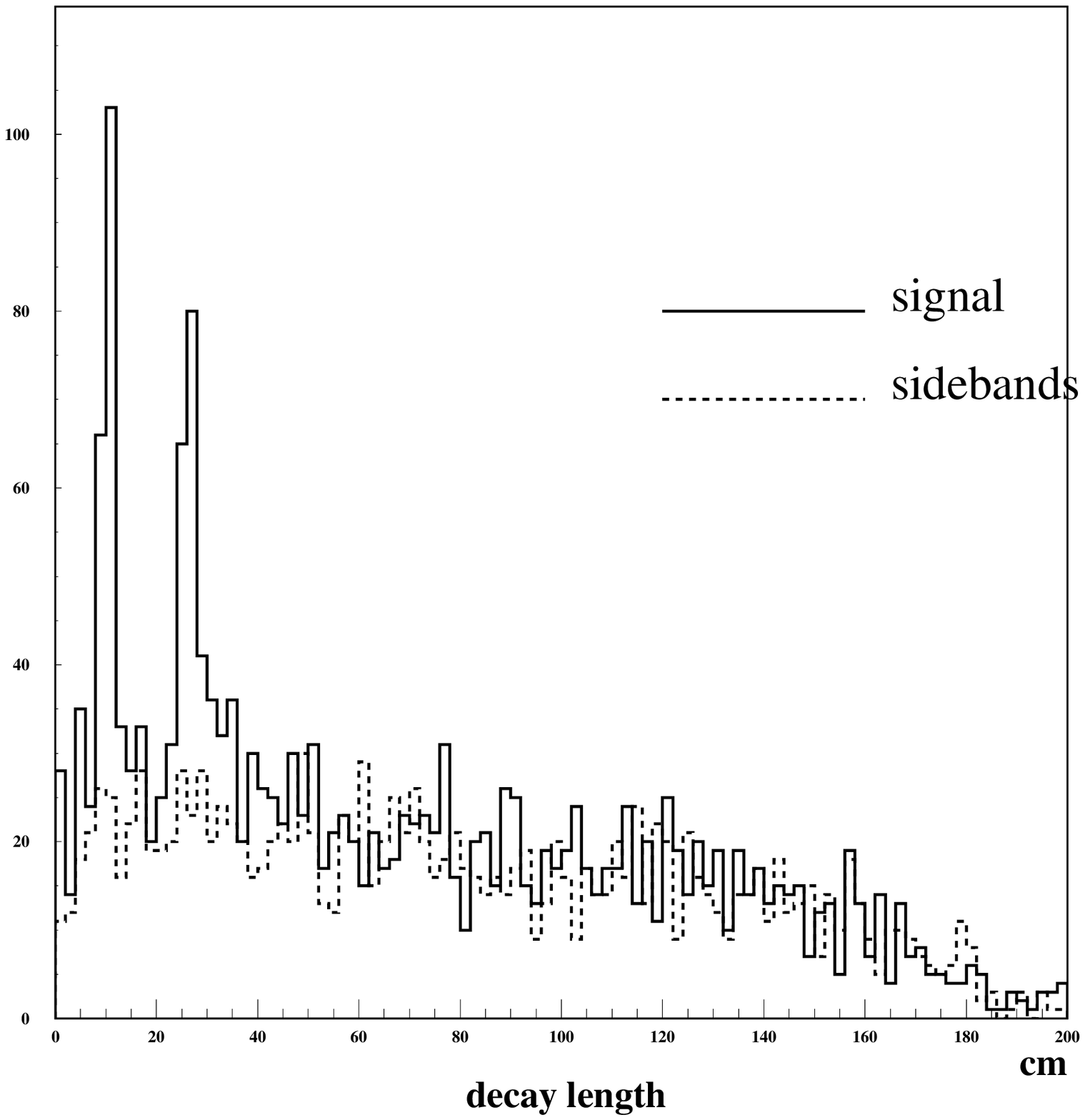,width=10cm}
\caption{Distribution of the decay distance for the signal and the side bands}
\label{KL2PDISTANCE}
\end{center}
\end{figure}

%\begin{figure}[htb]
%       \centering
%       \includegraphics[width = 3cm]{kloefigure.eps}
%       \caption{Distribution of the decay distance for the signal and the 
%       side bands}
%       \label{KL2PDISTANCE}
%\end{figure}

For both $K_{L} \to \pi^{+} \pi^{-}$ decays and 
$K_{L} \to K_{S} \to \pi^{+} \pi^{-}$ elastic events the absolute 
values of the $K_{S}$ and $K_{L}$ momenta should be equal. 
Fig.\ref{KSKLDELTAP} shows the distribution of the difference 
$\Delta|\vec{p}| = |\vec{p}_{S}| - |\vec{p}_{L}|$ 
for the signal band and the side-bands. 
The signal band was fitted with a gaussian centred at 
$\Delta|\vec{p}| = 0 $ MeV with $\sigma = 2$ MeV, 
representing the decays and a second gaussian centred at 
$\Delta|\vec{p}| = 3$ MeV with $\sigma = 4$ MeV, probably
representing elastic interactions where a small fraction of the momentum 
is transferred to the recoil nucleus.
To further reduce the background of semileptonic decays we required 
$-6 < \Delta|\vec{p}| < +12$ MeV. This cut selected 915
events. 

\begin{figure}
\begin{center}
\epsfig{file=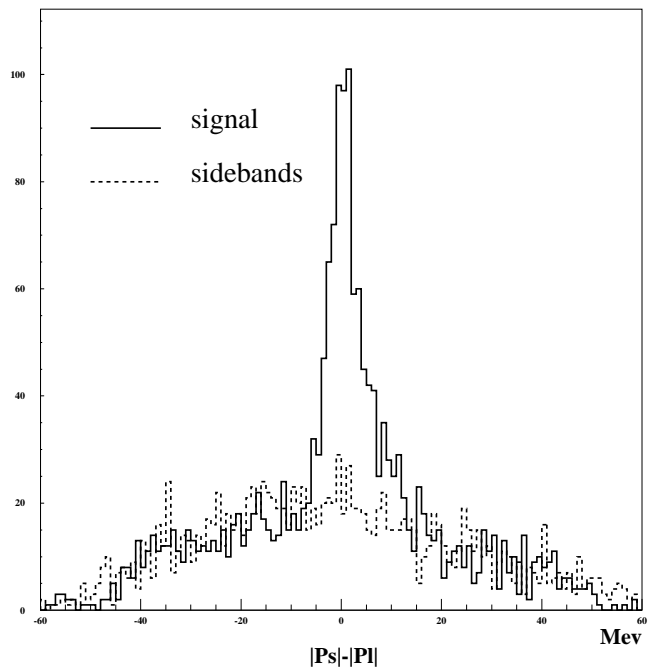,width=10cm}
\caption{Distribution of the difference between the absolute values of the $K_{S}$ and $K_{L}$ momenta for the signal and the side bands}
\label{KSKLDELTAP}
\end{center}
\end{figure}

%\begin{figure}[htb]
%       \centering
%       \includegraphics[width = 3cm]{kloefigure.eps}
%       \caption{Distribution of the difference between the absolute values 
%       of the $K_{S}$ and $K_{L}$ momenta for the signal band and the side 
%       bands}
%       \label{KSKLDELTAP}
%\end{figure}

Fig.\ref{KLPROJDISTANCE} shows the distribution of the vertex distance, $r$, 
and of the projected vertex distance, $\rho = (x^{2} + y^{2})^{1/2}$, for 
the 915 events and for the side-bands sample. On the basis of the $K_S$
decay length distribution of fig.\ref{lifefit}, 
two regions of interest were defined of width (-2,+4) cm around the
position of the regenerators: 
\begin{itemize}
\item the beam pipe region, 8 cm $< r <$ 14 cm, which contains 151 events
\item the chamber wall region, 23 cm $< \rho <$ 29 cm, which contains 156 events.
\end{itemize} 

We parametrized   
the $r$ and $\rho$ distributions of the decay vertex using two gaussians for the
peaks and a linear background, and we evaluated the amount of
background events in the regions of interest.
The number of $K_{L} \to K_{S} \to \pi^{+} \pi^{-}$ regenerated events is then: 
$N^{bp}_{reg} = 123 \pm 13$, $N^{cw}_{reg} = 122 \pm 12$.

\begin{figure}
\begin{center}
\epsfig{file=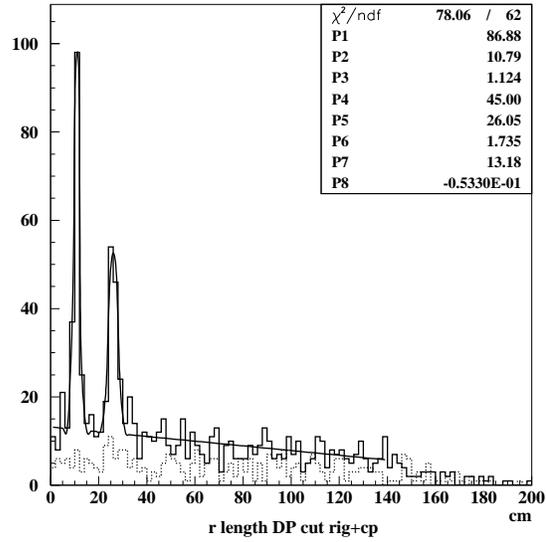,width=8cm}
\epsfig{file=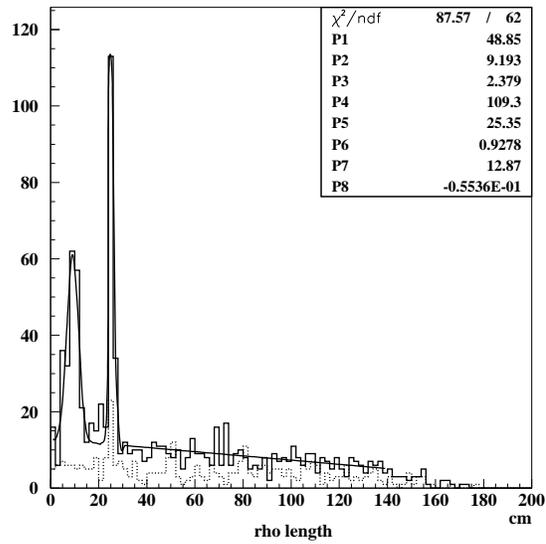,width=8cm}
\caption{Distribution of the decay distance in space and of its projection on the $x$-$y$ plane for $K_{L} \to \pi \pi$ decays and regenerated events, and for the side bands sample}
\label{KLPROJDISTANCE}
\end{center}
\end{figure}

%\begin{figure}[htb]
%       \centering
%       \includegraphics[width = 3cm]{kloefigure.eps}
%       \caption{Distribution of the decay distance in space and of its 
%       projection on the $x$-$y$ plane for $K_{L} \to \pi \pi$ decays and 
%       regenerated events, and for the side bands sample}
%       \label{KLPROJDISTANCE}
%\end{figure}

The distribution of the angle, $\omega$, between the $K_{S}$ and 
$K_{L}$ momentum, shown in Fig.\ref{KSKLDIRECTION}, 
was used to separate the 
$K_{L} \to \pi^{+} \pi^{-}$ decays, peaked at small angles, from the 
$K_{L}$ semileptonic decays and 
$K_{L} \to K_{S} \to \pi^{+} \pi^{-}$ elastic events. 
A cut at $\omega < 75$ mrad selected a sample of 279
$K_{L} \to \pi^{+} \pi^{-}$ decays.

\begin{figure}
\begin{center}
\epsfig{file=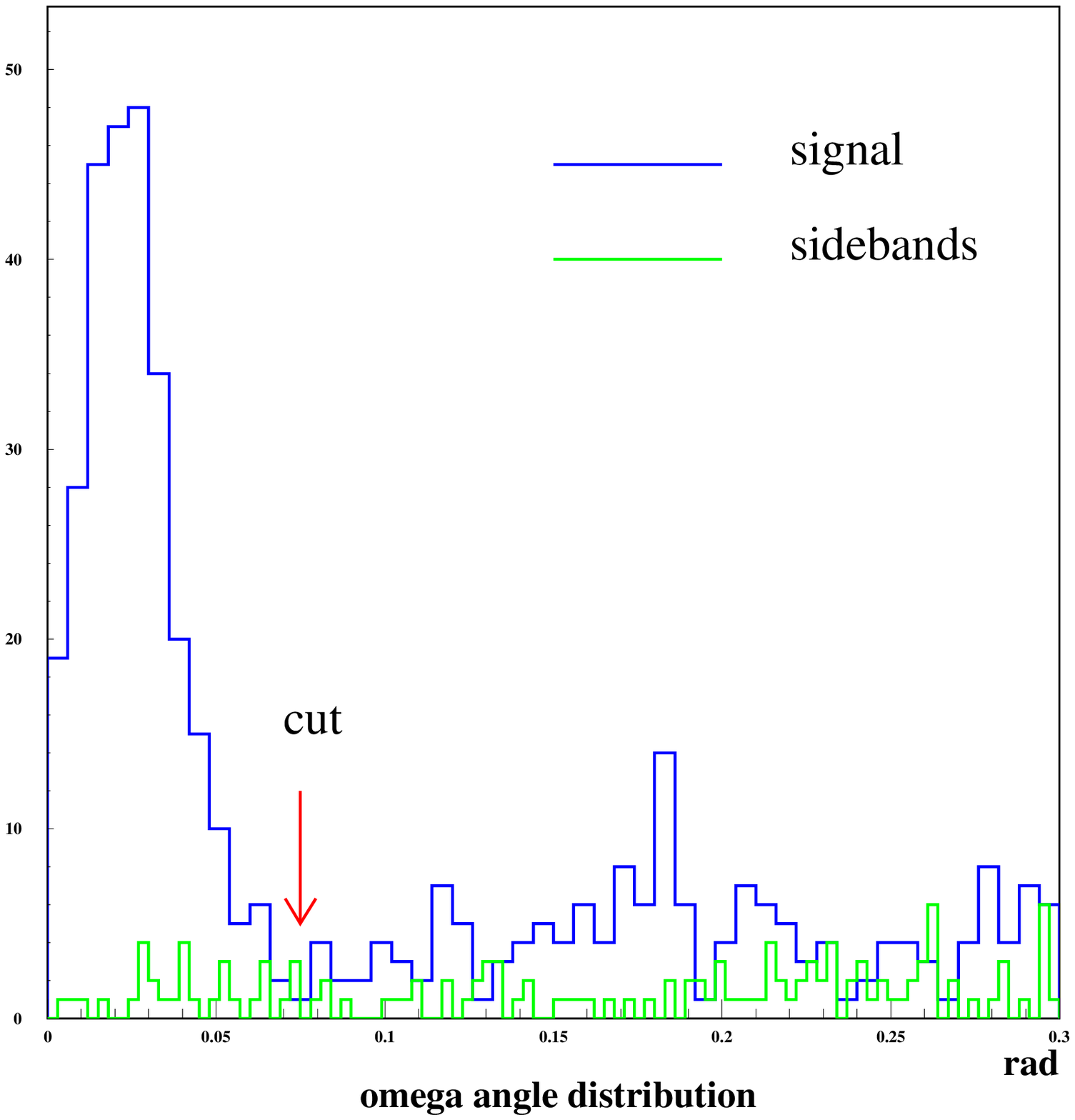,width=10cm}
\caption{Distribution of the angle between the $K_{S}$ and $K_{L}$ momenta for the signal and the side bands}
\label{KSKLDIRECTION}
\end{center}
\end{figure}

%\begin{figure}[htb]
%       \centering
%       \includegraphics[width = 3cm]{kloefigure.eps}
%       \caption{Distribution of the angle between the $K_{S}$ and $K_{L}$ momenta for the signal band and the side bands}
%       \label{KSKLDIRECTION}
%\end{figure}

\subsection{Results}

The number of $K_{L} \to K_{S} \to \pi^{+} \pi^{-}$ regenerated events 
is related to the regeneration cross section on nuclei, $\sigma_{reg}$, by
\begin{equation}
 N_{reg} = N_{L} \ \varepsilon \ B_{S \pi \pi} \ \sigma_{reg} \ n \ t 
\label{equ1}
\end{equation}
with $N_{L} = N_{SL} \ e^{- r/\lambda}$ the number of $K_{L}$ mesons reaching 
the regenerator, $\varepsilon$ the detection efficiency,
$B_{S \pi \pi}$ the $K_{S} \to \pi^{+} \pi^{-}$ branching fraction, 
$n$ the number of nuclei of the regenerator per unit volume, 
$t$ the thickness of the regenerator. 

The spherical beam pipe, of radius 10 cm,  is made of AlBeMet, an alloy of 61\%, in volume, of 
Beryllium ($\rho = 1.85$ g cm$^{-3}$) and 39\% of Aluminum 
($\rho = 2.7 $ g cm$^{-3}$) and has a thickness of 0.50 mm:
%\[ (nt)^{bp} = 6.0 \ 10^{23} \ 
%   \left( \frac{1.85}{9} \ 0.61 + \frac{2.7}{27} \ 0.39 \right) \ 0.05 =
%   4.93 \ 10^{21} \ cm^{-2} \]
\[ (nt)^{bp} =   4.93 \ 10^{21} \ cm^{-2} \]

The cylindrical drift chamber wall is made of Carbon 0.75 mm thick, 
60\%-fiber ($\rho = 1.72$ g cm$^{-3}$) and 40\%-epoxy ($\rho = 1.25$ g cm$^{-3}$) 
and has a 0.20 mm thick Aluminum shield:
%\[ (nt)^{cw} = 6.0 \ 10^{23} \ \left[
%   \left( \frac{1.72}{12} \ 0.6 + \frac{1.25}{12} \ 0.4 \right) \ 0.075 +
%   \frac{2.7}{27} \ 0.020 \right] = 6.97 \ 10^{21} \ cm^{-2} \]
\[ (nt)^{cw}  = 6.97 \ 10^{21} \ cm^{-2} \]

The crossing angle, computed event by event, gives an average increase of the 
thickness of 3$\%$ for the beam pipe and of 15.5$\%$ for the chamber wall.

The detection efficiency for $K_{L} \to K_{S} \to \pi^{+} \pi^{-}$ regenerated 
events is the same as for $K_{L} \to \pi^{+} \pi^{-}$ decays selected in the 
analysis, the only difference being the final $\omega$ cut. The 
efficiency is derived from the distribution of the 
$K_{L} \to \pi^{+} \pi^{-}$ events $dN_{L \pi \pi}/dr$, the 
decay length in the laboratory $\lambda$, and the number of 
$K_{S}K_{L}$ events
\begin{equation}
 \frac{N_{SL} \ B_{L \pi \pi}}{\lambda} \
e^{-r/\lambda} \ \varepsilon(r) \ = \frac{dN_{L \pi \pi}}{dr} 
\label{equ2}
\end{equation}
The experimental ${dN_{L \pi \pi}}/{dr}$ distribution for the 279 selected 
$K_L\rightarrow\pi^+\pi^-$ decays is shown in fig.\ref{cpradius}, where
also the linear fit used to extract the results is shown. 
The region below 4 cm, coinciding with the $K_S$ fiducial volume, has been
excluded from the fit to avoid ambiguities, while the region above 140 cm
has a drop in efficiency, being near the edge of the drift chamber volume.\par

\begin{figure}
\begin{center}
\epsfig{file=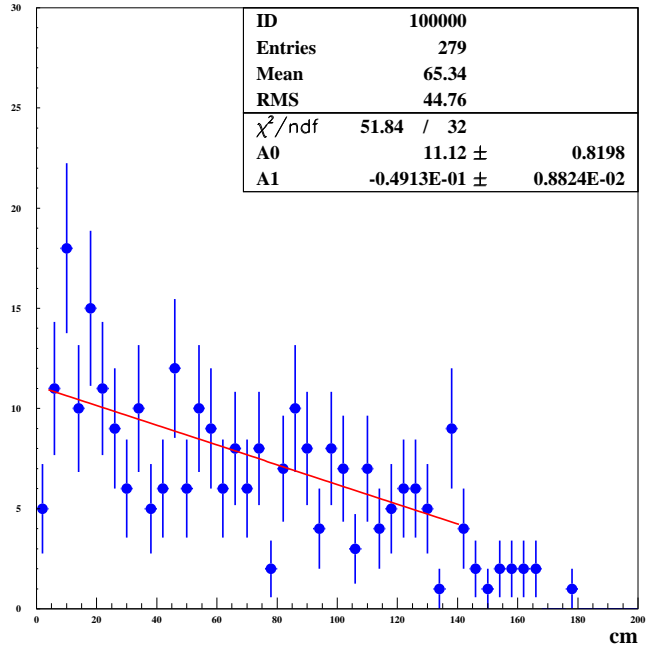,width=10cm}
\caption{Radial distribution of the CP violating $K_L\rightarrow \pi^+\pi^-$
decays, selected from data as explained in the text}
\label{cpradius}
\end{center}
\end{figure}

From the relations \ref{equ1} and \ref{equ2} we obtain
\begin{equation}
 \sigma_{reg} = \frac{B_{L \pi \pi}}{B_{S \pi \pi}} \
\frac{1}{\lambda \ dN_{L}/dr} \ \frac{N_{reg}}{\langle nt \rangle} 
\label{equ3}
\end{equation}

%Actually the average detection
%efficiency over the region of interest should be used, but it can be safely 
%approximated with the efficiency at the average radius of the same region. 
$(dN_{L \pi \pi}/dr)$ was evaluated at the average radius of each region of
interest, yielding:
$(dN_{L \pi \pi}/dr)^{bp} = 2.65 \ \pm \ 0.19 \ cm^{-1}$,
$(dN_{L \pi \pi}/dr)^{cw} = 2.41 \ \pm \ 0.15 \ cm^{-1}$.
We obtain

\[ \sigma_{reg}^{Be-Al} = 
   \left(75.7 \ \pm \ 9.6 _{\ stat} \right) \ mb \]
\[ \sigma_{reg}^{C-Al} = 
   \left(51.9 \ \pm \ 6.2 _{\ stat}  \right) \ mb \]
where both the statistical errors coming from the event counting and the
efficiency evaluation have been included.\par

The angular distribution for regenerated events in the beam pipe and in the 
chamber wall is shown in Fig.\ref{REGENANGLE}. To reduce the 
background only events in 10 cm $< r <$ 13 cm, for the beam pipe, and in  
24 cm $< \rho <$ 27 cm, for the drift chamber wall were used. The 
small contamination from $K_{L} \to \pi^{+} \pi^{-}$ events is fully 
contained in the first bin, as was checked in the selected $K_{L} \to
\pi^{+} \pi^{-}$ sample. The background from semileptonic decays is
negligible, according to montecarlo estimations.\par

\begin{figure}
\begin{center}
\epsfig{file=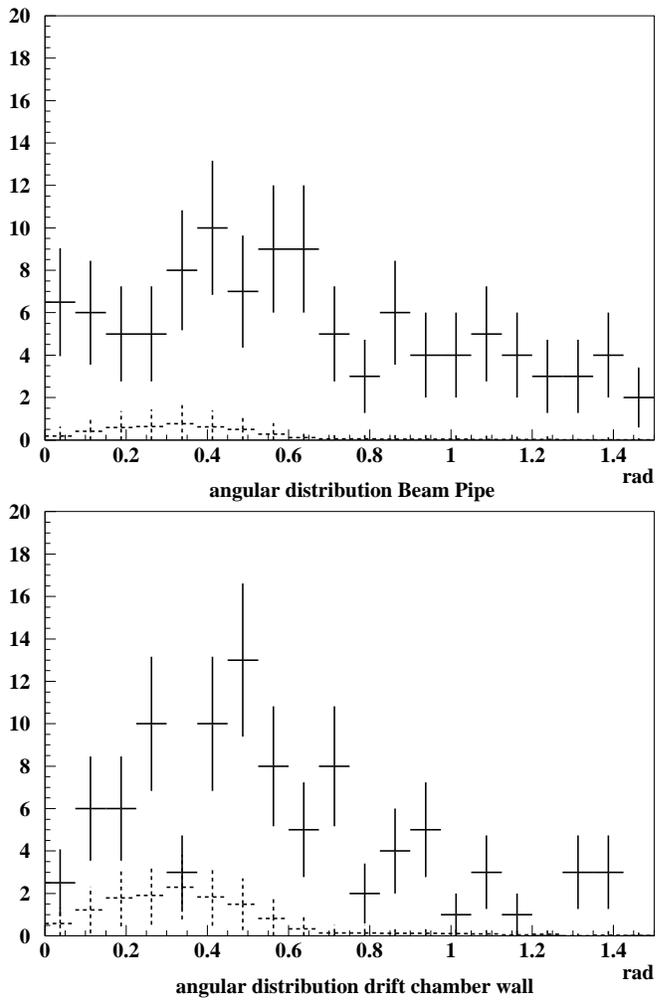,width=10cm}
\caption{Angular distribution of $K_{L} \to K_{S}$ regenerated events in the beam pipe and in the drift chamber wall}
\label{REGENANGLE}
\end{center}
\end{figure}

\subsection{Systematic error}
Three main categories of systematic error sources have been considered:
\begin{itemize}
\item event counting method;
\item efficiency evaluation;
\item regenerator thickness.
\end{itemize}
For the first category two sources of error have been studied: the definition
of the regions of interest (r.o.i) and the shape of the fit to the 
background. The r.o.i. limits were varied of $\pm$ 1 cm
obtaining maximum variations of about 3\% for the beam pipe events and of 
2\% for the drift chamber wall events. Various polynomial shapes were also
fitted to the background, obtaining 2$\div$3\% variations on the beam pipe
events, but much smaller (less than 1\%) on the drift chamber wall.\par
The efficiency evaluation is based on the assumption that regenerated
$K_S$ charged decays are detected with the same efficiency as the CP violating 
$K_{L} \to \pi^{+} \pi^{-}$ decays. A montecarlo study of such assumption
\cite{FERRARI} shows that differences in efficiency can be expected up to
2\%. 
%A cross check of this indication is obtained using the 
%``sidebands'' events which are almost completely given by semileptonic
%decays, according to montecarlo. Such decays are topologically more similar
%to regenerated decays (mainly looking at the $\omega$ angle distribution):
%the efficiency was then evaluated using their radial distribution
%(normalized to the number of $K_{L} \to \pi^{+} \pi^{-}$ selected decays),
%obtaining a variation below 2\% with respect to the standard analysis.\par

A possible contamination of regenerated $K_S$ decays in the $K_{L} \to
\pi^{+} \pi^{-}$ sample could bias the measurement to higher efficiencies.
To account for this effect the $K_{L} \to \pi^{+} \pi^{-}$ radial
distribution, showed in fig.\ref{cpradius}, was fitted adding to the main
linear shape two gaussians centered at the regenerators positions (just as
was done in fig.\ref{KLPROJDISTANCE}). The data are compatible with absence
of contamination on the drift chamber wall, but (probably due to a 
statistical fluctuation) allow a non negligible contamination on the beam
pipe, which results in a considerably reduced efficiency (7$\div$8\%). This
turns out to be the main systematic error (non related to the regenerators
knowledge) on this measurement.\par
Finally the beam pipe thickness is known with a precision of $50\,\mu
m$, according with the production tolerances.
For the drift chamber wall an uncertainty of $\pm 50\,\mu m$ on both carbon
and aluminum was estimated from the study of the multiple scattering of 
$e^+e^-$ pairs.

The systematic
error sources are summarized in table \ref{TABLE2}.\par

\begin{table}[htb]
\centering
\begin{tabular}{| c | c | c |}
\hline
 error source & $\Delta\sigma^{Be-Al}\, (mb)$ & $\Delta\sigma^{C-Al}\, (mb)$  \\
\hline\hline
 r.o.i. limits & 2.5 & 1.0 \\
 backgr. shape & 2.0 & 0.4 \\
\hline
 $\varepsilon_{cp}\neq\varepsilon_{reg}$ & 1.5  & 1.1 \\
 regen. contam. in $K_{L} \to \pi^{+} \pi^{-}$ & 6.5 & 3.7 \\
\hline
 regenerators thickness & 7.6 & 3.5 \\
\hline\hline
total & 10.6 & 5.3 \\
\hline
\end{tabular}
\caption{Systematic error contributions.}
\protect\label{TABLE2}
\end{table}

%\begin{figure}[htb]
%       \centering
%       \includegraphics[width = 3cm]{kloefigure.eps}
%       \caption{Angular distribution of $K_{L} \to K_{S}$ regenerated events 
%       in the beam pipe and in the drift chamber wall}
%       \label{REGENANGLE}
%\end{figure}

\section{Conclusions and discussion}
On the basis of the first data collected during the commissioning of
DA$\Phi$NE the KLOe experiment has analyzed $\sim 1.4\, 10^5$
$\phi\rightarrow K_LK_S$ decays with both kaons decaying to charged
particles.
Using Bhabha scattering events to precisely define the collision region, 
and fitting the distribution of the $K_S\to\pi^+\pi^-$ decay vertex, the
$K_S$ decay length is measured as:
$$ \lambda_s\, =\, 5.78\, \pm\, 0.08\, (stat)\, \pm\, 0.10\, (syst)\, mm.$$

The average $K_L$ decay lenght is $\lambda_L = 333 \pm 13\, (stat)$ cm.
A sample of 279 $K_{L} \to \pi^{+} \pi^{-}$ $CP$-violating decays 
is clearly identified with negligible background, and is used to measure
the efficiency for 
reconstructing $K_{L} \to K_{S} \to \pi^{+} \pi^{-}$ events due to regeneration
in the beam pipe and in the drift chamber inner wall.

The cross section in the beam pipe made of a 
61\%Beryllium-39\%Aluminun alloy is
$$\sigma^{Be-Al} = 75.7 \pm 9.6_{stat} \pm 10.6_{syst} \, mb$$ 
The regeneration cross section in the drift chamber wall made mainly of Carbon is
$$\sigma^{C-Al} = 51.9 \pm 6.2_{stat}\pm 5.3_{syst} \, mb$$
With this data
it is not possible to evaluate separately the regeneration cross section on 
Beryllium, Carbon and Aluminun nuclei unless we make additional 
hypotheses on its dependence upon the atomic mass. Theoretical 
calculations \cite{BALDINI} based on the eikonal approximation 
are shown in 
Fig.\ref{CROSSBALDINI} as well as a measurement on Beryllium
ref.\cite{SOLODOV1} \cite{SOLODOV2} made with the CDM-2 detector 
at the Novosibirsk VEPP-2M electron-positron collider.
The theoretical predictions do not show any evident dependence upon 
the atomic mass due to the interference of the $K$ and $\bar{K}$
amplitudes that have different behaviour as a function of $A$. 
However 
our measurements can hardly accomodate a cross section on Alluminium nuclei 
smaller than for Beryllium and Carbon, as shown in
fig.\ref{COMP} where the correlation between the values of the cross sections
is drawn.\par

\begin{figure}
\begin{center}
\epsfig{file=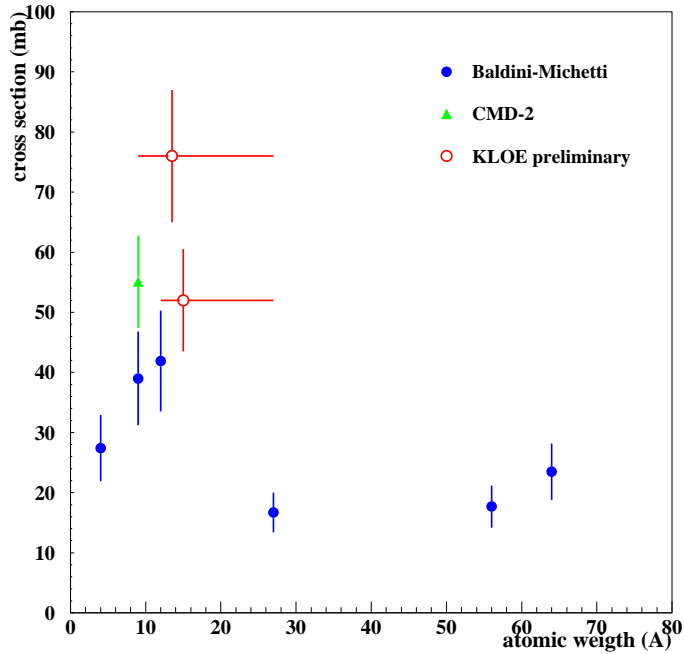,width=10cm}
\caption{$K_{L} \to K_{S}$ regeneration cross section as a function of the atomic mass.}
\label{CROSSBALDINI}
\end{center}
\end{figure}

\begin{figure}
\begin{center}
\epsfig{file=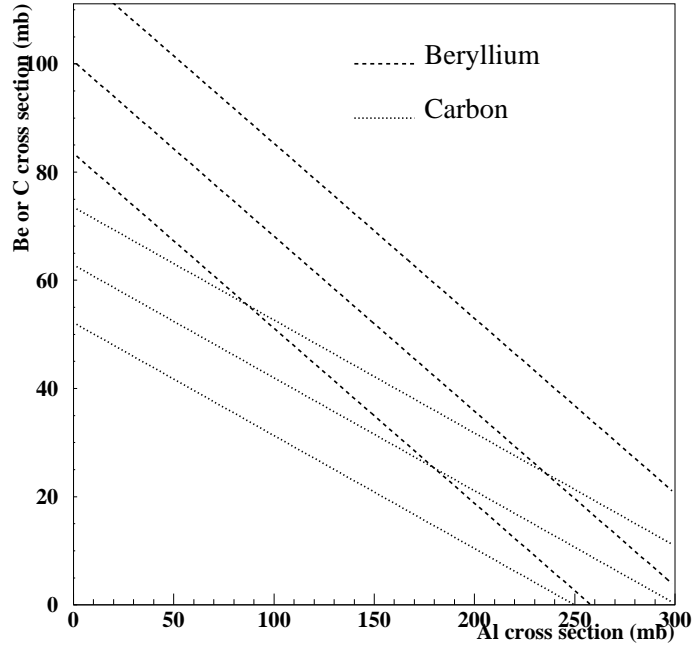,width=10cm}
\caption{Beryllium and Carbon cross section as a function of the Alluminum regeneration cross section.}
\label{COMP}
\end{center}
\end{figure}

%\begin{figure}[htb]
%       \centering
%       \includegraphics[width = 3cm]{kloefigure.eps}
%       \caption{Experimental results as a function 
%       of the atomic mass.}
%       \label{CROSSBALDINI}
%\end{figure}


\begin{thebibliography}{99}

\bibitem{DAFNE}S.Guiducci et al., \textit{DA$\Phi$NE Operating Experience},
  Proceedings of PAC99, New York, March 1999.


\bibitem{BERTOLUCCI} S.Bertolucci, \textit{A Status Report of KLOE at 
 DA$\Phi$NE}, Proceedings of the 19th International Conference on 
 Lepton and Photon Interactions at High Energy, Stanford 1999.


\bibitem{EVCL} C.Bloise,M.Incagli,\textit{The event classification module
    and the ECLO,  ECLS banks}, KLOE memo 175, 1/99.

\bibitem{DAF99} \textit{Status of the KLOE Detector}, Proc. of DA$\Phi$NE99, Frascati (1999).

\bibitem{TRKMON} S.Dell'Agnello,\textit{TRKMON, The Tracking Monitor
    Program of KLOE}, KLOE Memo 202, 1/99.

\bibitem{PAIS} A.Pais and O.Piccioni, Physical Review 100 (1955) 1487

\bibitem{KLEINKNECHT} K.Kleinknecht, Fortschritte f\"ur Physik 21 (1973) 57

\bibitem{BALDINI} R.Baldini and A.Michetti, 
 \textit{$K_{L}$ interactions and $K_{S}$ regeneration in KLOE}, 
 LNF-96-008 (IR), 16.2.1996. 
 
 
\bibitem{PALUTAN} M.Palutan, \textit{CP-violation study at 
DA$\Phi$NE: the trigger of the KLOE experiment}, Ph.D. Thesis, University 
of Rome ``Tor Vergata'', 1999.

\bibitem{FERRARI} A.Ferrari, \textit{Measurement of the $K_L\to K_S$
    regeneration cross section in the KLOe experiment at DA$\Phi$NE}, Ph.D.
Thesis, University of Rome ``La Sapienza'', 1998.


\bibitem{SOLODOV1} E.P.Solodov, 
 \textit{Study of the rare $K_{S}$, $K_{L}$, $K^{+}$, $K^{-}$ decays at the 
 $\phi$ resonance with the CMD-2 detector}, 
 Proceedings of the 29th International Conference on High Energy Physics, 
 A.Astbury, D.Axen and J.Robinson eds., World Scientific 1999, p.1046.
 
\bibitem{SOLODOV2} E.P.Solodov, 
 \textit{Studies of kaon decays at the $\phi$ resonance}, Proceedings of
 the 1999 Workshop on Kaon Physics, eds. J.L.Rosner and B.R. Winstein, The
 University of Chicago Press, p.311.
 
\end{thebibliography}
\end{document}